**A Reinforcement Learning Approach for GNSS Spoofing Attack Detection of Autonomous Vehicles**

**Sagar Dasgupta***
**Ph.D. Student**
Department of Civil, Construction & Environmental Engineering
The University of Alabama
3014 Cyber Hall Box 870205, Tuscaloosa, AL 35487
Tel: (864) 624-6210; Email: sdasgupta@crimson.ua.edu

**Tonmoy Ghosh**
**Ph.D. Student**
Department of Electrical and Computer Engineering
The University of Alabama, Tuscaloosa
3067 South Engineering Research Center, Tuscaloosa, AL 35487
Tel: (205) 657-8375; Email: tghosh@crimson.ua.edu

**Mizanur Rahman, Ph.D.**
**Assistant Professor**
Department of Civil, Construction & Environmental Engineering
The University of Alabama
3015 Cyber Hall, Box 870205, Tuscaloosa, AL 35487
Tel: (205) 348-1717; Email: mizan.rahman@ua.edu

*Corresponding author

Word count: 5,133 words text + 4 table x 250 words (each) = 6,133 words

Submission date: August 1, 2021

Paper submitted for presentation at the Transportation Research Board 101st Annual Meeting and for publication in Transportation Research Record

**ABSTRACT**

A resilient and robust positioning, navigation, and timing (PNT) system is a necessity for the navigation of autonomous vehicles (AVs). Global Navigation Satelite System (GNSS) provides satellite-based PNT services. However, a spoofer can temper an authentic GNSS signal and could transmit wrong position information to an AV. Therefore, a GNSS must have the capability of real-time detection and feedback-correction of spoofing attacks related to PNT receivers, whereby it will help the end-user (autonomous vehicle in this case) to navigate safely if it falls into any compromises. This paper aims to develop a deep reinforcement learning (RL)-based turn-by-turn spoofing attack detection using low-cost in-vehicle sensor data. We have utilized Honda Driving Dataset to create attack and non-attack datasets, develop a deep RL model, and evaluate the performance of the RL-based attack detection model. We find that the accuracy of the RL model ranges from 99.99% to 100%, and the recall value is 100%. However, the precision ranges from 93.44% to 100%, and the f1 score ranges from 96.61% to 100%. Overall, the analyses reveal that the RL model is effective in turn-by-turn spoofing attack detection.

**Keywords:** Reinforcement Learning, Cybersecurity, GNSS, GPS, Autonomous vehicle, and Spoofing attack

## INTRODUCTION

With the advancement of communication and automation technologies, the landscape of roadway mobility systems changes radically *(1)*. Ground vehicles are becoming more automated as well as connected between themselves and with the transportation infrastructure. It provides a traveler an opportunity to efficiently move from one location to another location and use their time for personal use while traveling. Positioning, navigation, and timing (PNT) services are the key to the navigation of autonomous vehicles (AVs) *(2)*. Global Navigation Satelite System (GNSS) provides satellite-based PNT services. In the United States, GNSS is known as the global positing system (GPS) *(3)*. GPS is a set of satellites that the US Department of Defense launched in 1970 for military use. However, GPS is known initially as Navigation Satellite Timing and Ranging (NAVSTAR). Such as GPS, several countries have their own satellites for providing PNT services. In the US, GPS provides two different services— i.e., Standard Location Service (SPS) and Precision Location Service (PPS) *(3)*. As PPS service is available for government and military use, it is expected that autonomous vehicles will use SPS, which is also known as civilian GPS, for their PNT services.

AVs require reliable and real-time PNT services. However, the robustness and reliability of GNSS-based PNT services depend on strong satellite signals and radio communications at the receiver end. The long distance between satellites and GNSS receivers reduces the signal strength and decreases GNSS-based PNT services' reliability. The GNSS signal is also susceptible to natural vulnerabilities, which are known as unintentional vulnerabilities *(4)*. For example, a GNSS signal can be unavailable to an autonomous vehicle while passing through a tunnel. Even ceilings in garages and thick clouds in the sky could reduce the GNSS signal strength and interrupt the PNT services. In urban areas, tall buildings cause multipath propagation, which causes radio frequency interference and degrades GNSS signal strength *(4)*. Besides these vulnerabilities, jamming and spoofing are the common intentional threats to GNSS-based PNT services *(2),(4)*. A jamming attack makes the authentic GNSS signal unavailable to a receiver by flooding a high-power compromised signal. On the other hand, a spoofing attacker can temper an authentic GNSS signal and could transmit wrong position information to an AV. Even an AV's destination and route choice can be corrupted, and the vehicle could be misguided turn-by-turn to an unwanted destination. This will compromise the safety of AV users. Nowadays, an attacker can use low-cost software-defined radios to conduct such a spoofing attack.

Many methods exist to detect a sophisticated GNSS spoofing attack by analyzing GNSS signals *(5)-(9)*. Encryption mechanisms, codeless-cross-correlation measures, signal statistics analysis, and antenna-based approaches are common techniques for detecting spoofing attacks. In addition, researchers also use in-vehicle inertial navigation system (INS) and inertial measurement units (IMU) sensors, such as gyroscope and accelerometer, to calculate a vehicle's position and acceleration to flag a spoofing attack by comparing with GNSS-based position and acceleration information. However, these in-vehicle sensors provide less reliable position, speed, and orientation information because of scale factor and non-orthogonality errors *(10)(11)*. In addition, these errors accumulate over time. With the advancement of machine learning (ML) and deep learning (DL) models, several studies investigated the potential of sophisticated spoofing attack detection using ML and DL *(12)(13)*. However, to the best of the authors' knowledge, there is no study that uses a deep reinforcement learning (RL) approach to detect a sophisticated turn-by-turn spoofing attack.

This paper develops a deep RL approach using data from multiple low-cost in-vehicle sensors of an AV to detect a sophisticated turn-by-turn spoofing attack. The presented RL approach will be a new addition to the existing GNSS spoofing attack detection approaches without directly analyzing the GNSS signal characteristics and avoid IMU/INS-based solutions because of their inability to provide accurate position and acceleration information.

## RELATED WORK

Although many studies exist related to spoofing attack detection of GNSS using encryption mechanism, codeless-cross-correlation measures, signal statistics analysis, and antenna-based strategy *(4), (5)-(9),* we have reviewed existing literature related to machine learning and in-vehicle sensors because of their relevance to this study's focus.

Borhani-Darian et al. used the Cross Ambiguity Function (CAF) feature to develop a deep learning approach for detecting spoofing attacks *(14)*. For the model development purpose, they have combined a multi-layer perceptron (MLP) and two classes of convolution neural networks (CNNs), which include a complex CNN and a simple CNN, to provide a probability-based attack classification. The deep learning approach was trained and tested using simulated data and proved the potential for detecting spoofed GNSS signals. In another study, Sun et al. *(12)* used singular values of the wavelet transformation feature for the spoofed and actual GPS signal to train three different attack classifiers: (i) support vector machines (SVM); (ii) probabilistic neural networks (PNN); and (iii) decision tree (DT). Later, they fused the individual classification outcome of these three models with a K-out-of-N rule, which increases the detection accuracy on average by 3.75%, 5.06%, and 12.36% compared to the SVM, PNN, and DT itself, respectively. In addition, their K-out-of-N decision rule showed a fewer number of false-positive than each of those three classifiers. Panice et al. presented a GPS spoofing attack detection approach for Unmanned Aerial Vehicle (UAV) using SVM *(13)*. They have used SVM to estimate the state of a UAV and identify anomalies of its current location. This approach constructs a decision boundary through training using the data from the actual state of the UAV. The uniqueness of this approach is that it provides a probability if it misses any detection, which is necessary for aviation applications. However, if a spoofer has complete knowledge of a UAV's trajectory data, their detection system could not detect the attack. Instead, it gives high position errors. On the other hand, Shafiee et al. presented a new MLP-NN approach for GPS spoofing attack detection *(15)*. They have selected three features from the GPS signal pattern—i.e., early-late phase criterion, delta criterion, and total levels of signal— to train and test the performance of their MLP-NN approach. They found 98.7% accuracy for detecting spoofed signals, and computation time is less than 0.5 seconds. They also compared the performance of K-Nearest neighbor (KNN) and naive Bayesian classifier with the MLP-NN to prove the efficacy of their method. It was revealed that MLP-NN provides a high accuracy compared to the other three classifiers. All of these above-mentioned deep learning and machine learning based attack detection strategies in the GNSS signal level. In addition, vehicle position information has not been used to detect GNSS spoofing attacks.

On the other hand, several studies presented inertial navigation system (INS) and inertial measurement units (IMU) sensor-based spoofing attack detection strategies as these sensors are resilient against signal spoofing attacks and provide a low-cost solution. Gyroscope and accelerometer are two examples of INS and IMU sensors. A spoofing attack can be detected by comparing IMU-based acceleration and GNSS-based acceleration *(16)*. However, this approach is not suitable for autonomous ground vehicles as they have a low vehicle dynamics signature compared to the aircraft. Manickam and O'Keefe have compared position information from the accelerometer and gyroscope with the position information from GNSS to flag a spoofing attack *(17)*. They also analyzed different types of IMU combining with different GNSS receiver grades to evaluate their performance for attack detection. Researchers also used INS to keep track of the position of a vehicle, and eventually, it helps to detect GNSS spoofing attacks *(11),(12)*. Even only the dead reckoning approach has been used to determine the speed, orientation, and position of a vehicle using gyroscope and accelerometer data. However, INS-based position, speed, and orientation information are less reliable because of scale factor and non-orthogonality errors, which increase over time *(18)*.

In this study, we present an RL-based spoofing attack detection strategy using multiple low-cost in-vehicle sensors' data of an autonomous ground vehicle, which has not been explored previously. The contribution of this study is the development of the RL strategy using a real-world Honda Driving Dataset.

## DATASET DESCRIPTION AND DATA PROCESSING

This study uses a real-world driving dataset –i.e., Honda Research Institute Driving Dataset (Honda dataset) *(19)*, to develop a RL-based GPS spoofing attack detection model. The Honda Dataset includes different vehicle sensor data for suburban and urban driving scenarios, which makes it more appropriate to be used for developing a spoofing attack detection model. Specifically, this dataset is suitable for training AV GPS spoofing detection model for the urban environment in which the AV is more vulnerable. It is a challenge to detect the shift in location in urban scenario because of a compromised GPS information. The dataset contains data of 104 hours of real human driving in the San Francisco Bay area with a vehicle equipped with AV sensors. **Figure 1** presents a sample driving route of the vehicle**.**

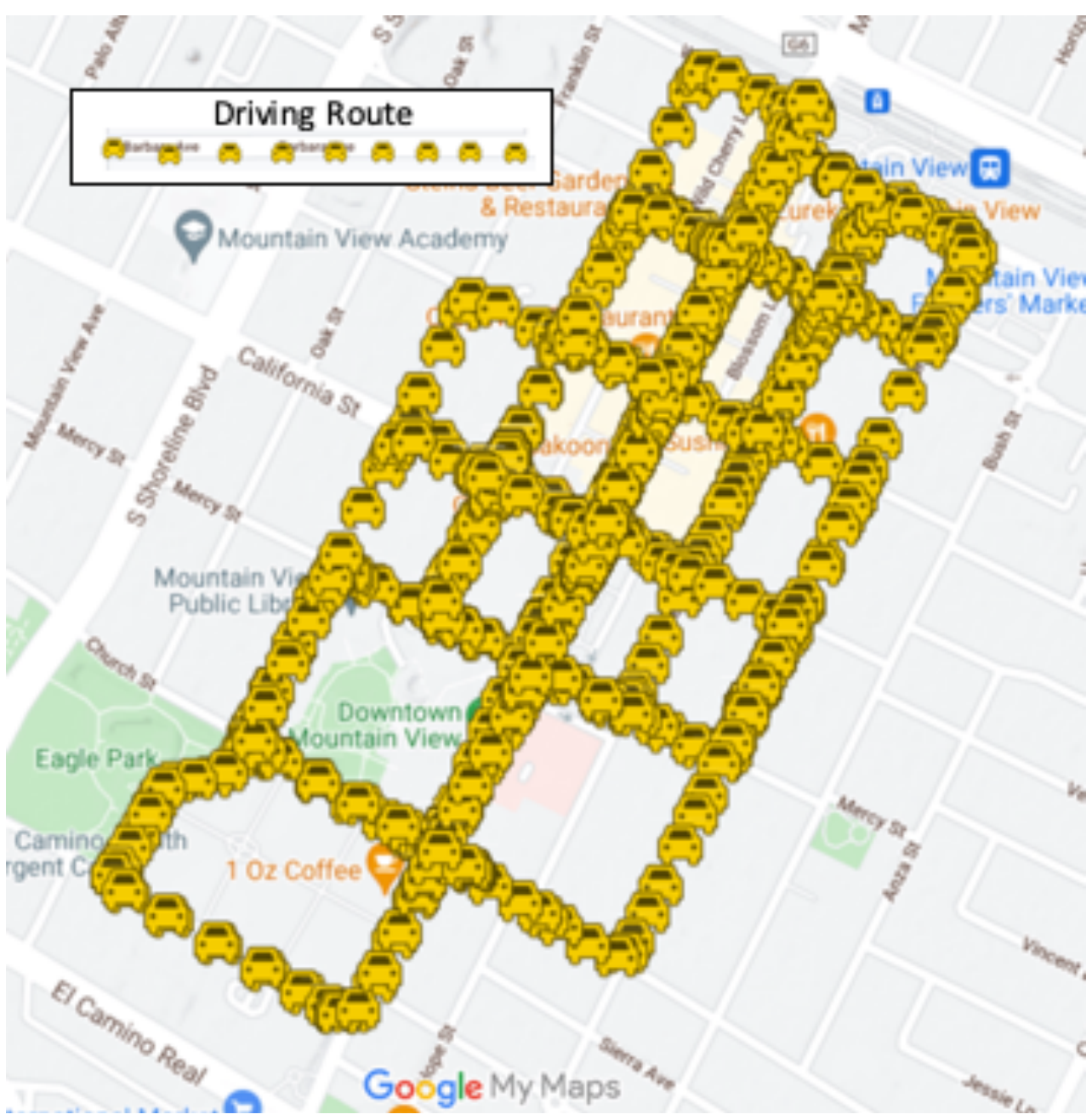

**Figure 1. Sample driving routes from Honda Research Institute Driving Dataset**

The vehicle was equipped with camera, LiDAR, GPS, inertial measurement unit (IMU), and control area network (CAN) sensors. GeneSys Eletrinik GmbH Automotive Dynamic Motion Analyzer with DGPS is used for sensing GPS, accelerometer, and gyroscope data at 120 Hz. However, the gyroscope and accelerometer data are not included in the Honda Dataset. The CAN output includes the throttle angle, brake pressure, steering angle, yaw rate, and speed sampled at 100 Hz.

This study uses latitude and longitude information from GPS and steering wheel angle (deg), speed (ft/s), and relative accelerator pedal position (%) from CAN to create a dataset for training and testing a deep RL model for detecting GPS spoofing attacks. The sampling rate of GPS and CAN output is not the same; therefore, the raw data are synchronized, keeping the GPS Unix timestamp as a reference. The GPS latitude and longitude information are used to calculate the distance traveled between two consecutive timestamps using the Haversine formula *(20),* as shown in **Equation 1**:

$$d = 2r\, sin^{-1}\left(\sqrt{sin^2\left(\frac{\varphi_2 - \varphi_1}{2}\right) + cos(\varphi_1)\, cos(\varphi_2)\, sin^2\left(\frac{\psi_2 - \psi_1}{2}\right)}\right) \qquad (1)$$

where d is the distance in meters between two points on the Earth's surface; r is the earth's radius (6378 km); $\varphi_1$ and $\varphi_2$ are the latitudes in radians; $\psi_1$ and $\psi_2$ are the longitudes in radians of two consecutive time stamps. **Table 1** presents a sample raw sensor data from Honda Dataset. In addition, **Figure 2** presents sensor data that we have used for creating the training and testing dataset to develop a deep RL model.

**TABLE 1 Sample Data from Honda Dataset**

| Unix Timestamp | GPS | | Speed (ft/s) | Steering Wheel Angle (deg) | Accelerator Pedal Position (%) |
|---|---|---|---|---|---|
| | Latitude | Longitude | | | |
| 1488224209.42714 | 37.393 | -122.077 | 0 | -57.8 | 0 |
| 1488224209.43716 | 37.3939 | -122.077 | 0 | -57.8 | 0 |
| 1488224209.44696 | 37.3939 | -122.077 | 0 | -57.8 | 0 |
| 1488224209.45698 | 37.3939 | -122.077 | 0 | -57.8 | 0 |
| 1488224209.46698 | 37.3939 | -122.077 | 0 | -57.8 | 0 |

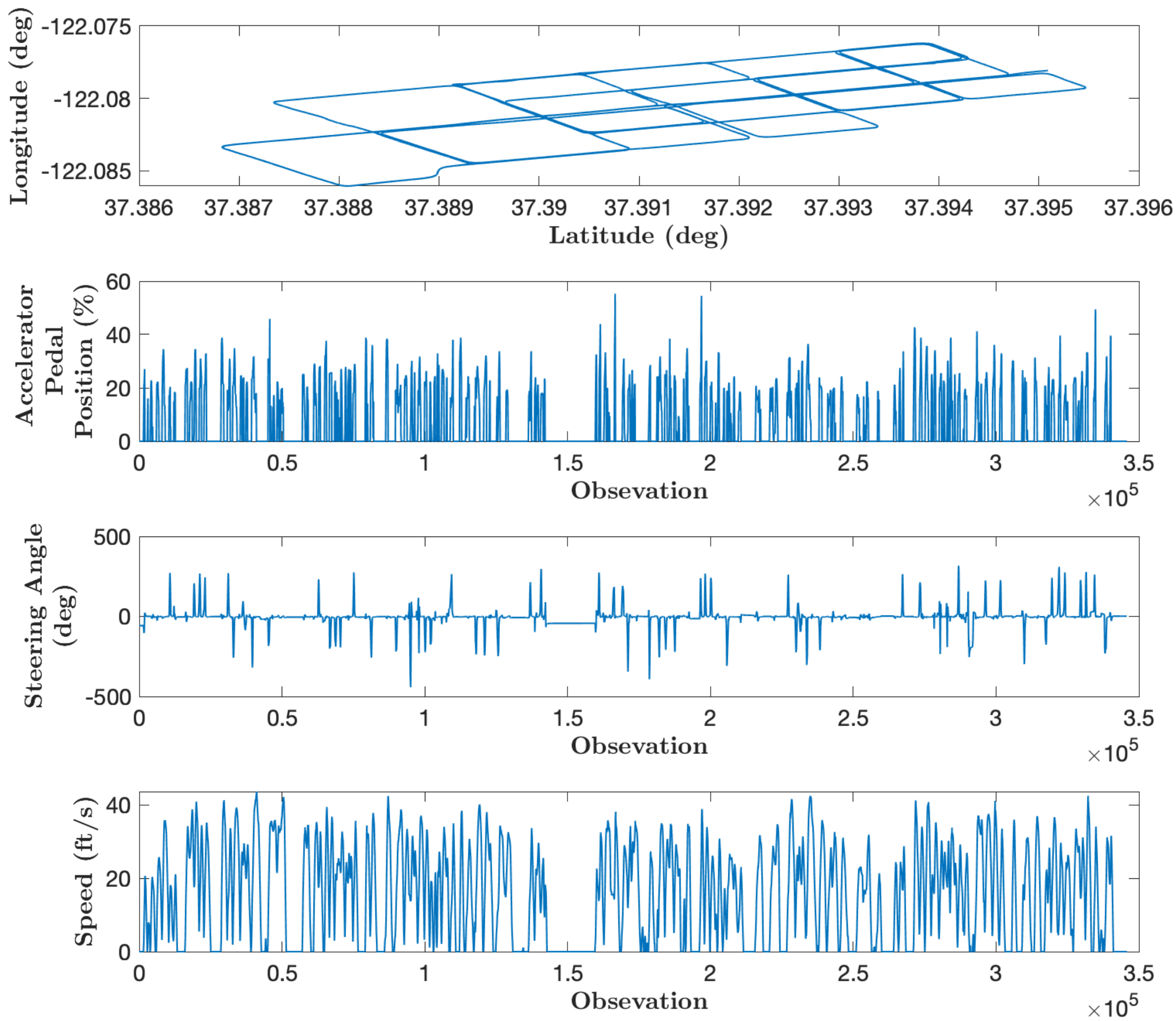

**Figure 2 GPS and CAN data from Honda Dataset**

## ATTACK MODEL

In this study, we have created a turn-by-turn *(21)* spoofing attack, which is a practical and sophisticated GPS spoofing attack for AV navigation. In order to create such an attack, a spoofer requires the destination information of a target AV. During this attack, a spoofer generates wrong GPS signal and makes the AV GPS receiver lock onto the spoofed signal. An AV believes the spoofed signal as an authenticate GPS signal. After taking control over the GPS signal, a spoofer creates a spoofed route matching all the turns of the actual route. Thus, it is a challenge to detect such anomalies. An example of such an attack is shown in **Figure 3.** Here, the blue-colored line is the suggested route created at the beginning of a trip based on the origin and destination information by an AV user. When the AV reaches location A, the spoofer takes over the AV GPS, and it shifts the location from A to B. Due to the change of current location, the navigation application creates another route from B to destination, whereas the AV's actual location is A. Now, the spoofer keeps updating the AV's location in such a way that vehicle will believe that it is moving along the black-dashed route; however, the AV is moving along the orange route. Both the black and orange routes have the same number of turns and the same type (right or left) of turns. As a result, the AV ends up at a spoofed destination without recognizing that it is under attack.

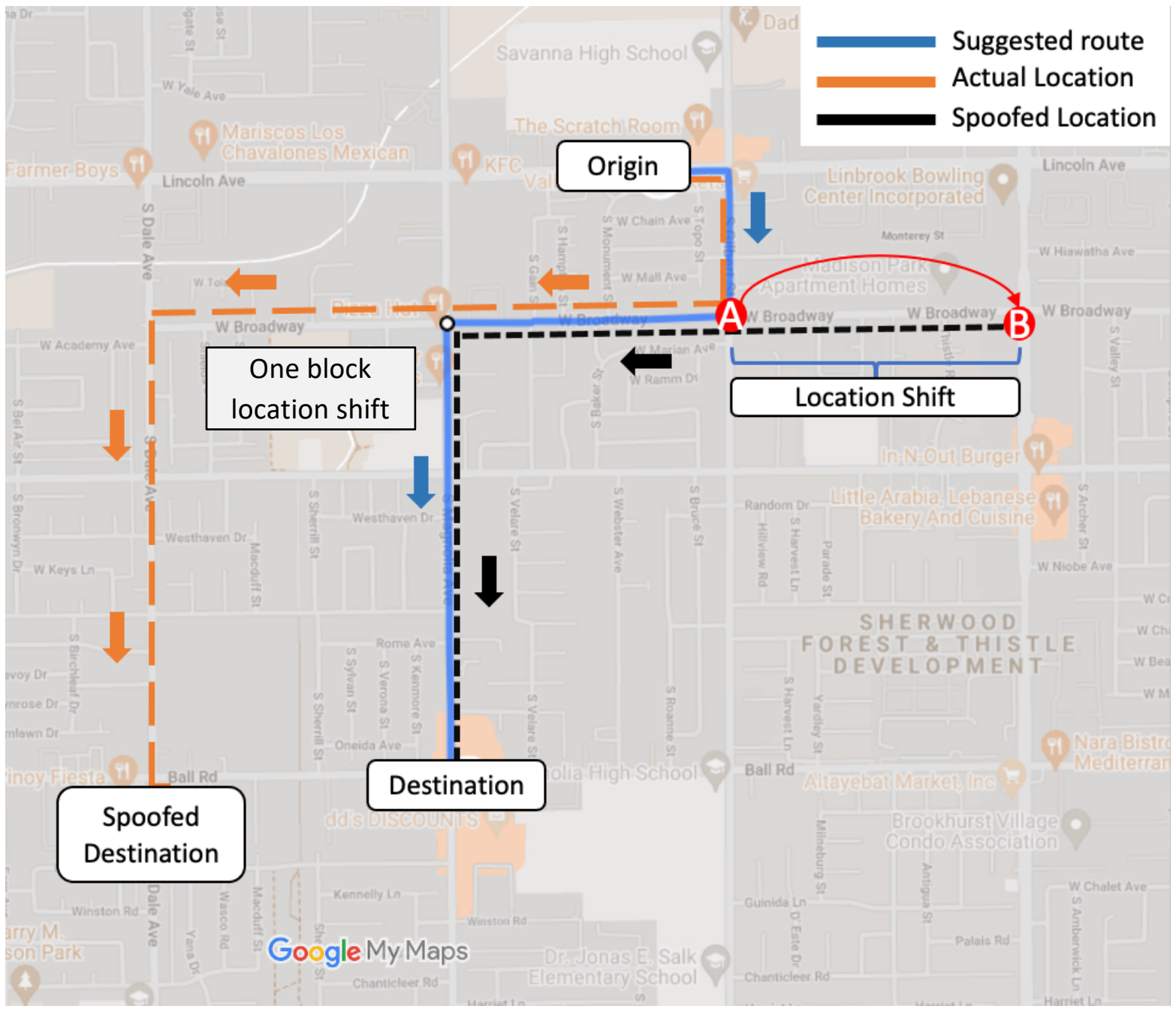

**Figure 3 Turn-by-turn GPS spoofing attack**

## GENERATION OF TURN-BY-TURN ATTACK DATASET

We have created attack-free and attack dataset for developing a GPS spoofing attack detection model. A total of ten different attack datasets are created to train and test data the RL model. Note that all attack datasets include multiple turn-by-turn attacks. One of the basic features of the turn-by-turn attack is the location shift. In our datasets, the location shift ranges from one block to a couple of blocks shift of location (See **Figure 3**). We have also created the location shift at different locations, which ranges from 50 m to 180m. **Figure 4** presents location shift values for all ten datasets. For the attack-free scenarios, the distance traveled between two consecutive timestamps is much lower than the location shift. All the picks in **Figure 4** represent the attack locations where a location shift occurs. Scenario 1 contains the highest number of attacks, and scenario 10 contain the least amount of attacks.

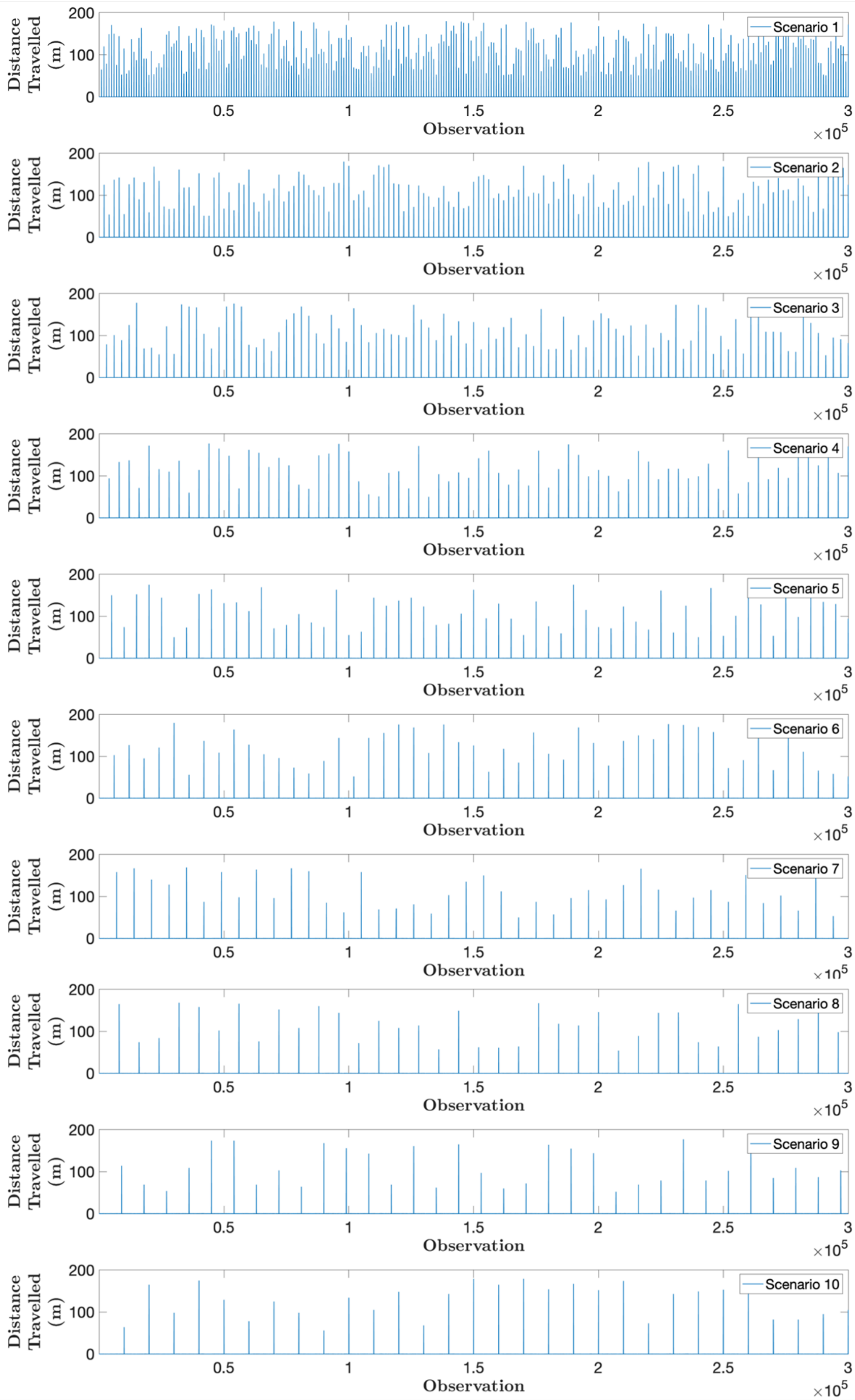

**Figure 4 Attack free and attack datsets**

## REINFORCEMENT LEARNING BASED DETECTION MODEL DEVELOPMENT

In this study, we develop an RL-based turn-by-turn spoofing attack detection framework, as shown in **Figure 5(a)**. The threshold value obtained from the trained RL model is compared with the differential distance (DD), which is calculated using real-time GNSS data from an AV. The absolute difference between the predicted and calculated distance is defined as DD. If real-time DD is greater than the threshold, then an attack is detected; otherwise, no attack is detected. **Figure 5(b)** presents an RL model, which consists of two components: (i) agent and (ii) environment *(22).* We have defined the RL problem inside the environment. Here, AV sensor data (GPS and CAN) are used to calculate and predict the distance traveled by an AV between two consecutive timestamps. The ground truth GPS trust status—i.e., whether the GPS is compromised or not— and the DD is then fed to the agent. The agent will adjust the threshold value and compare it with the differential distance so that the reward is maximum. Note that if a GPS is compromised, the DD must be higher than the adjusted threshold value.

An agent is designed to observe the environment and trained so that it behaves optimally in a given environment state, which results in a partial or complete solution. To achieve the optimal solution, an agent interact with the environment in discrete time steps. An agent's primary purpose is to choose actions that will maximize the overall future reward. The agent is always modifying its policy in order to discover the optimal one. The proposed RL method is represented in **Figure 5** as a flow diagram. In this study, we use the "keras-rl" reinforcement learning framework, which is built on keras, as a base. Note that, TensorFlow backend is used in keras.

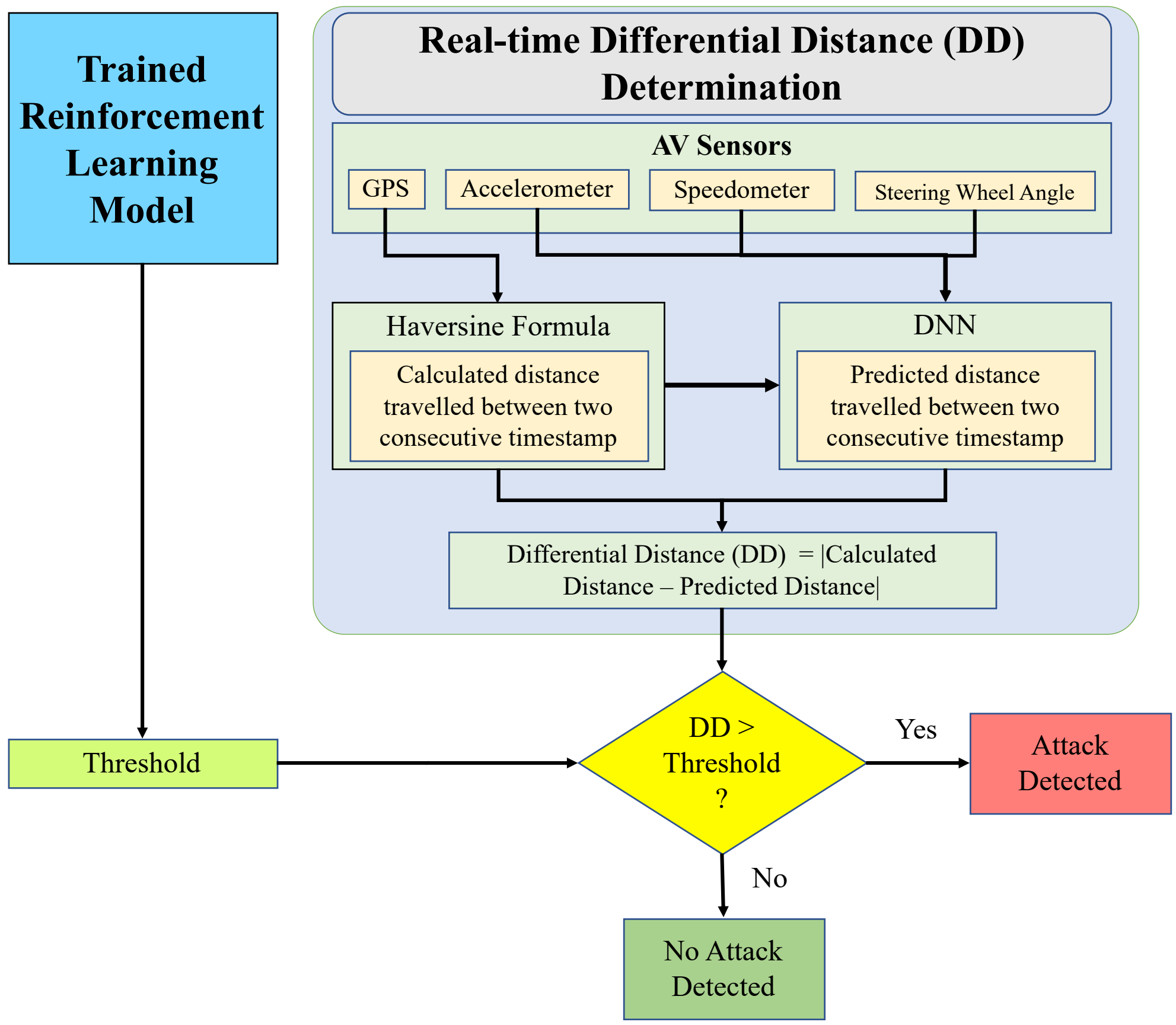

(a) Detection framework

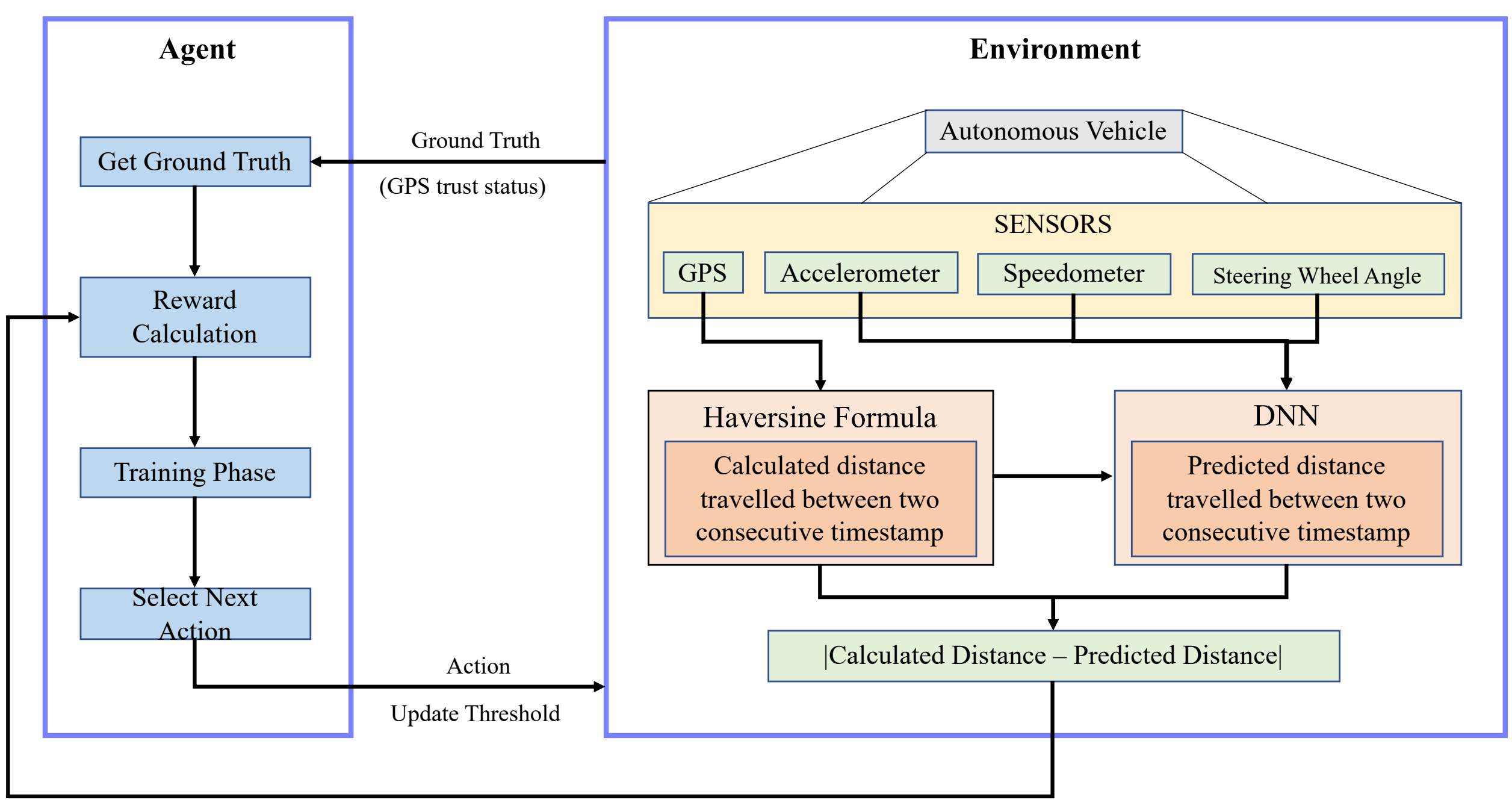

(b) Reinforcement learning model

**Figure 5 Deep RL-based turn-by-turn spoofing attack detection framework**

**Environment**

In our deep RL framework, the environment consists of an AV equipped with various sensors. The input sensor data consists of latitude, longitude, speed, steering wheel angle, and relative acceleration pedal position (%). We consider that the AVs GPS receiver is compromised during the spoofing attack; however, CAN output—i.e., speed, steering wheel angle, and relative acceleration pedal position— are not affected by the attacker. Distance traveled by an AV between two consecutive timestamps is calculated using the GPS coordinates as formulated in **Equation 1**. We have also predicted the distance traveled by the AV using a deep neural network (DNN) using the CAN sensor data. The training data also include the current GPS trust status—i.e., whether the GPS is compromised or not. The DNN predicts the distance travelled by an AV between two consecutive timestamps using CAN and GPS data.

DNN is an artificial neural network (ANN). It consists of multiple layers of interconnected single or multiple neurons in an input layer and an output layer. **Figure 6** depicts the DNN architecture, which is used in this study. The input layer has 4 neurons corresponding to three input data from CAN and one from GPS, which is distance traveled between two consecutive timestamps. There are three fully interconnected hidden layers with the number of neurons of 16, 8, and 4. The output layer has one neuron, and the output data is the predicted distance traveled between two consecutive time stamps by an AV. A rectified linear unit (ReLU) is used as the activation function for hidden layers. The DNN model is trained and validated using uncompromised (Honda Dataset) acceleration, steering wheel angle, speed, and GPS coordinate based distance traveled data. These raw sensor data are normalized between 0 and 1 prior to training. We have used 241,989 (70%) observations for training and 103,709 (30%) observations for validation. The DNN model hyperparameters—i.e., number of layers, inner layers width, inner layers neurons, epochs, optimizer, loss function, and activation function—for inner layers are selected by a trial-and-error approach. **Table 1** provides optimal hyperparameters values of the DNN model. To find a set of optimal hyperparameters, we have used Mean Absolute Error (MAE) metric for the loss function, which helps to identify if there are any model underfitting and overfitting issues. After validating the DNN model, we find that Root Mean Square Error (RMSE) and maximum absolute prediction error for the DNN are 1.18 x $10^{-5}$ m and 0.07m, respectively. **Figure 7(a)** presents a comparison between the ground truth and predicted distance traveled (i.e., location shift) for the validation result, and **Figure 7(b)** shows the absolute error profile between the ground truth and the predicted data.

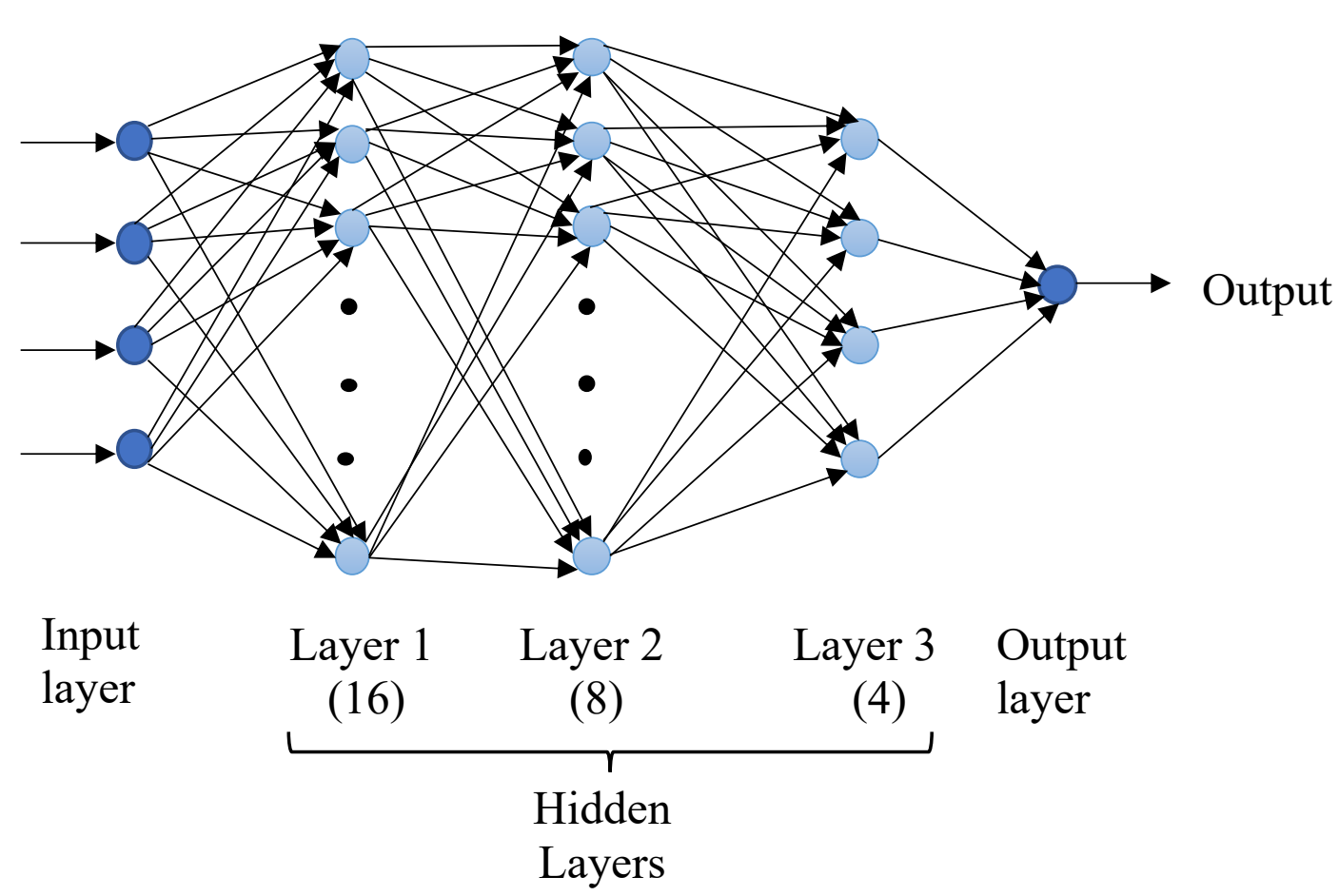

**Figure 6. DNN-based distance traveled prediction model architecture**

**TABLE 2 DNN Model Hyperparameters**

| Hyperparameters | Value |
|---|---|
| Number of layers | 5 |
| Inner layers width | 3 |
| Number of neurons each layer | 4, 16, 8, 4, 1 |
| Number of epochs | 1000 |
| Optimizer | ADAM |
| Loss function | Mean Average Error |
| Activation function | ReLU |

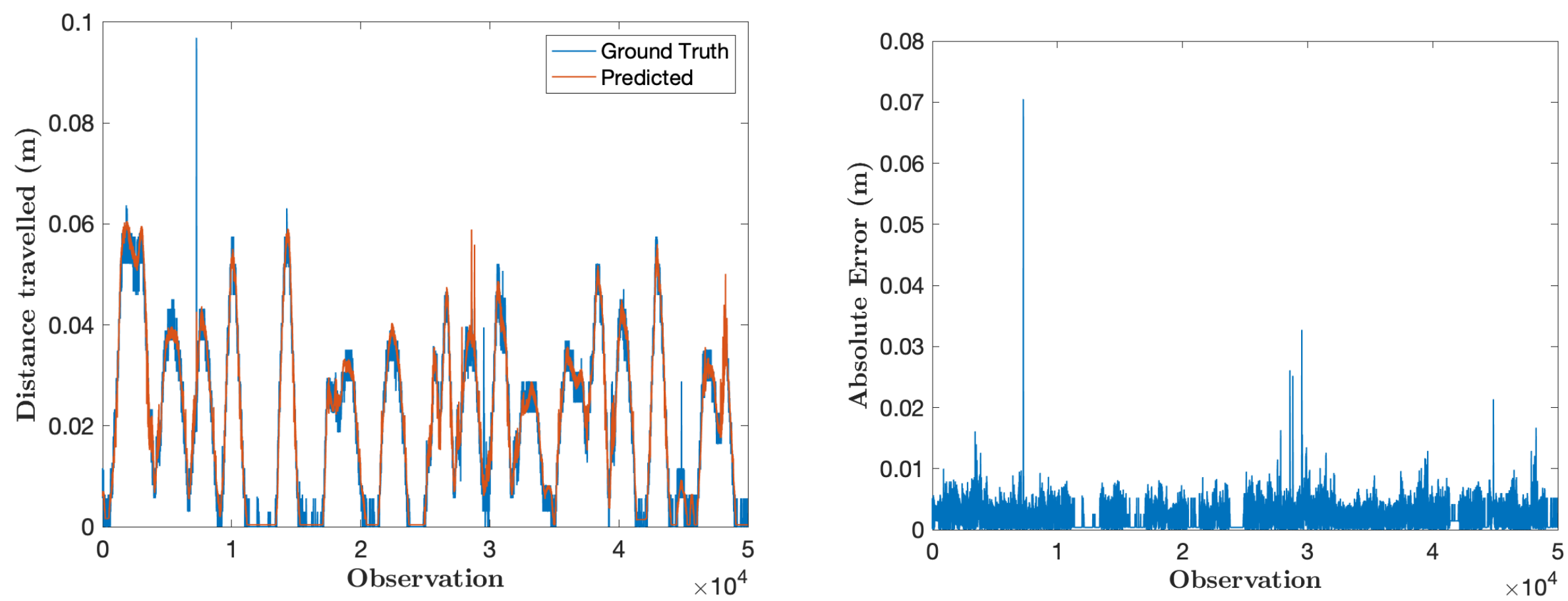

(a) (b)

**Figure 7 (a) Comparison of ground truth data and predicted data; and (b) Absolute error profile**

## Agent

We have presented a RL architecture comprised of a single agent system. Deep Q-Networks (DQN) agent in "keras-rl" is used as an agent. There are four steps inside the agent: (i) get ground truth; (ii) calculate reward; (iii) training phase; and (iv) select next action. After obtaining the ground truth GPS trust status and the differential distance from the environment, an agent calculates the reward function. However, an agent acts based on an optimal policy. The state of this agent is a threshold value, which is defined based on the maximum prediction error for distance traveled between two consecutive timestamps. If the assigned threshold value is lower than the differential distance, it detects that the GPS is compromised. Later, the agent will also check with the ground truth data; if the ground truth matches with the agent detection, a positive reward (+1) is given. On the other hand, if an agent's detection outcome doesn't match with the ground truth, then a higher negative reward (-100) is given. We assign a much higher negative reward to prioritize detecting an attack.

## Deep Q-learning Algorithm

We have used a Deep Q-learning algorithm for the deep RL framework. Q in Q-learning represents quality and learning represents an objective to choose a policy that will maximize the total reward. The Q-learning function takes random actions outside the current policy and learns the detection policy. For this reason, Q-learning is called an off-policy RL algorithm. It also creates a Q-learning table Q[s,a] (it represents the current state (s) of the environment) and corresponding rewards for each possible action (a), which is Q-value. The algorithm chooses an action (a) with the highest reward, which is called Q-score. The Q-value is defined as formulated in **Equation 2.**

$$Q(s_t, a_t) = Q(s_t, a_t) + \alpha(r_{t+1} + \gamma\, maxQ(s_{t+1}, a) - Q(s_t, a_t)) \quad (2)$$

where, $s_t$ is the time state, $a_t$ is the action state, $\alpha$ is the learning rate, $r$ is the reward, and $\gamma$ is the discount factor. Therefore, we have calculated the current Q-values based on the current and next states and actions, learning rate, and discount factor. The objective of the algorithm is to optimize the total reward intake and Q-learning by using the experience from current and next states and actions. The discount factor can take a value between 0 and 1, which regulates the importance of the immediate and future reward.

**Figure 8** presents the details of our deep RL framework. Here, a DNN model is also used to approximate the Q-learning function. The input for training the DNN is the predicted DD obtained from the environment. This DNN architecture consists of four layers. The first layer is the input layer, and then there are two fully interconnected hidden layers with 24 neurons in each layer. The last layer is the output layer, which has three neurons. The input neuron accepts the differential distance, and the output layer can give one of the three following outputs: (i) increase threshold; (ii) decrease threshold; and (iii) keep the threshold the same. As described before, we have also used the same "keras-rl" reinforcement learning framework. TensorFlow backend is used in keras, and DQN agent is used in keras-rl. Only data for scenario 2 from the attack dataset is used for the model training. The DNN model parameters are listed in **Table 3.**

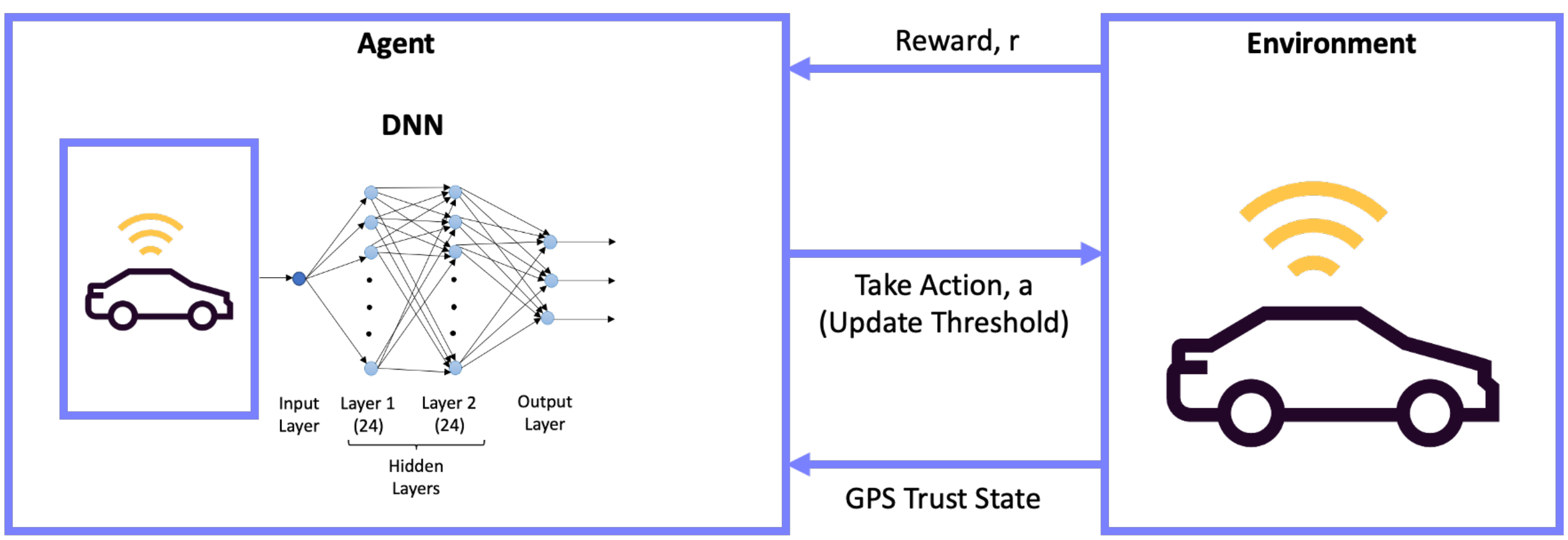

**Figure 8. Deep Q-learning Algorithm**

**Table 3. RL Model Hyperparameters**

| **Hyperparameters** | **Value** |
|---|---|
| Number of layers | 4 |
| Inner layers width | 2 |
| Number of epochs | 10000 |
| Optimizer | ADAM |
| Loss function | Mean Average Error |
| Activation function | ReLU |

# RESULTS AND DISCUSSION

We have evaluated the performance of deep RL framework accuracy, precision, recall and f1 score metrics using nine test datasets (scenarios 1, 3, 4, 5, 6, 7, 8, 9, 10; see **Figure 4**). The accuracy, precision, and recall are calculated using the confusion matrix generated from each model considering equal class weights. The precision-recall curve for the developed RL model is presented in **Figure 9**, which presents the model performance. A high precision value shows fewer false positives, and a high recall value proves fewer false negatives. So, we can conclude that our RL-based spoofing attack detection model performs well as it shows high precision and recall value. **Figure 10** presents the plots for all detection results using testing datasets. Here, X-axis represents the observation number, and Y-axis represents the state. State 0 means there is no attack and state 1 means there is an attack. Only the first 30,000 observations of the ground truth state and model predicted state is shown in **Figure 10.** It shows that the RL model can successfully detect an attack as soon as the attack is created—i.e., location shift occurred.

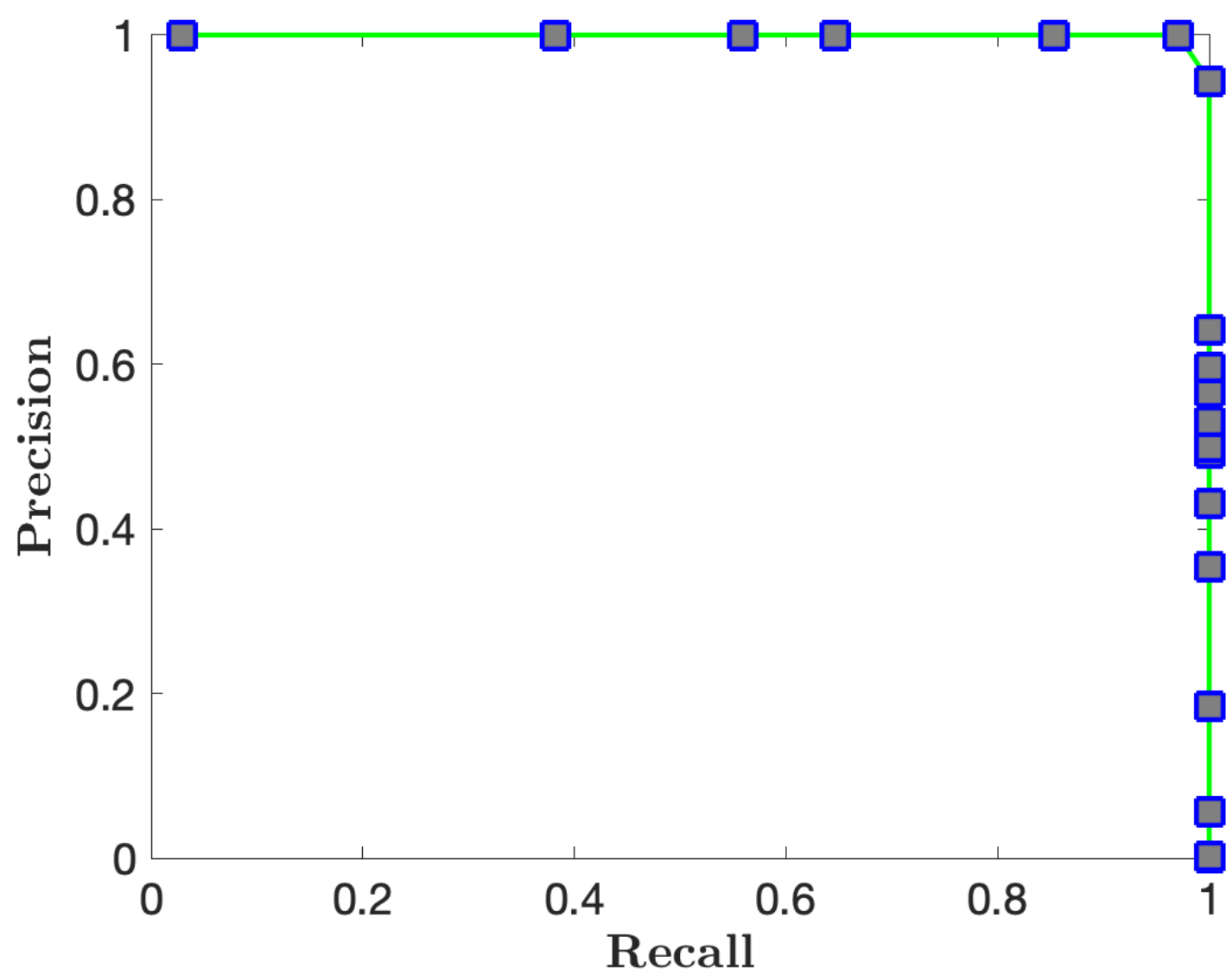

**Figure 9 Precision-recall curve for the testing datasets**

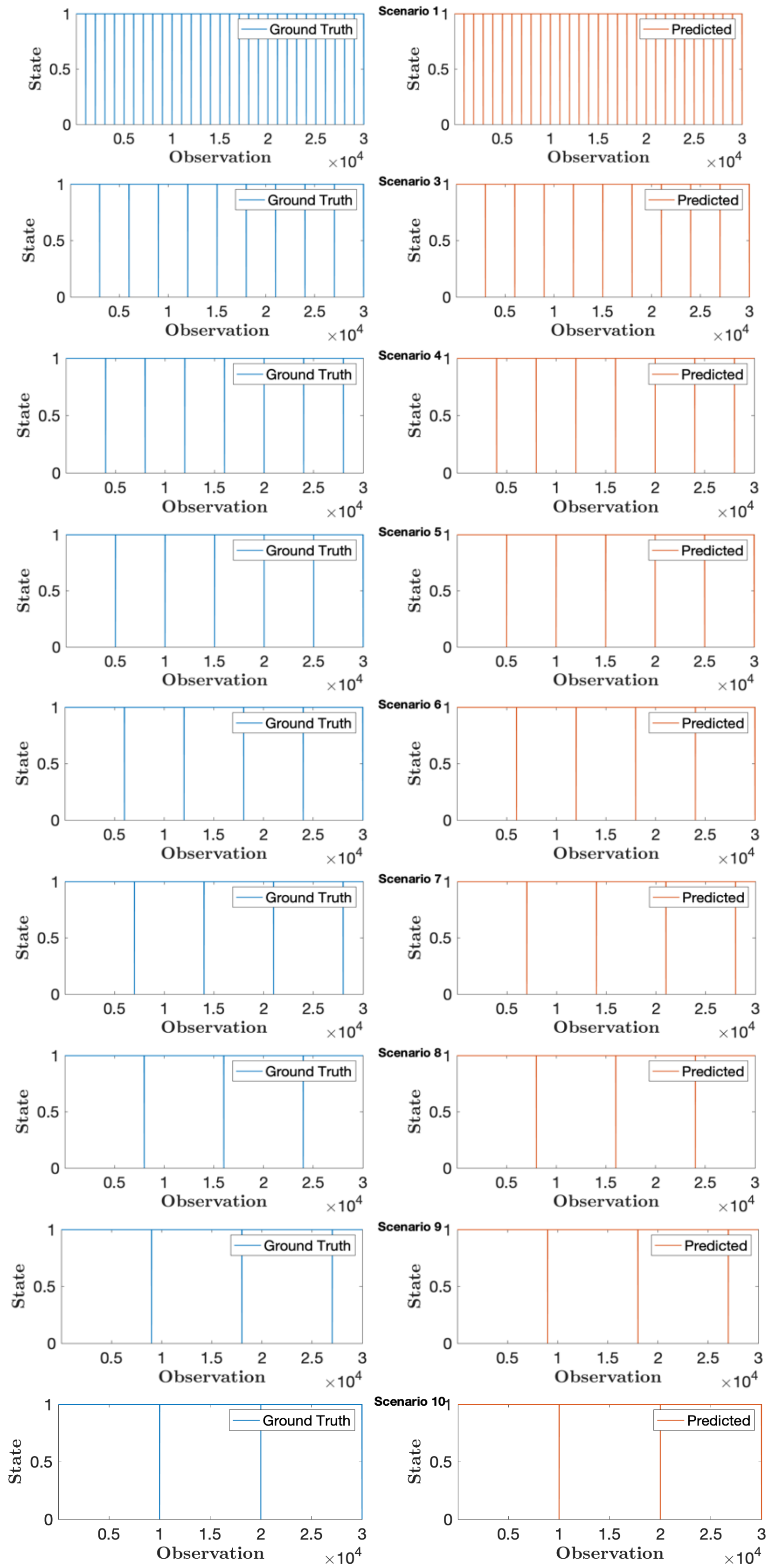

**Figure 10. A sample detection result (only for first 30,000 observations)**

**Table 4** provides the recall, precision, accuracy, and f1-score for nine test datasets, which represents nine scenarios. The recall value for all the test cases is 100%, which indicates that the developed model can detect all the attacks. The precision values range from 93.44% (scenario 6) to 100%. 100% precision value means that no false attack was detected. Precision less than 100% means there are instances where the model detected an attack incorrectly—i.e., there is no attack, and an attack is detected. The model achieved the lowest accuracy of 99.99% for seven attack scenarios, and it achieved the highest accuracy of 100% for 2 scenarios. The f1-score ranges from 96.61% to 100%, which means the false positives and false negatives are very low. Based on the performance for all the attack scenarios, we conclude that the deep RL model is effective in detecting GPS turn-by-turn spoofing attacks.

**TABLE 4 Model Evaluation Results**

| Attack Scenario | Recall | Precision | Accuracy | f1-score |
|---|---|---|---|---|
| 1 | 100% | 98.57% | 99.99% | 99.28% |
| 3 | 100% | 98.29% | 99.99% | 99.14% |
| 4 | 100% | 100% | 100% | 100% |
| 5 | 100% | 98.57% | 99.99% | 99.28% |
| 6 | 100% | 93.44% | 99.99% | 96.61% |
| 7 | 100% | 100% | 100% | 100% |
| 8 | 100% | 97.72% | 99.99% | 98.85% |
| 9 | 100% | 97.43% | 99.99% | 98.70% |
| 10 | 100% | 94.44% | 99.99% | 97.14% |

## CONCLUSIONS

A GNSS must have the capability of real-time detection and feedback-correction of spoofing attacks related to PNT receivers, whereby it will help the AVuser to navigate safely if it falls into any compromises. This paper developed an RL-based turn-by-turn spoofing attack detection model using low-cost in-vehicle sensor data. We have utilized Honda Dataset to create ten attack and non-attack datasets, develop the RL model, and evaluate the performance of the RL model for attack detection. We found that the accuracy of the RL model ranges from 99.99% to 100%, and the recall value was 100%. However, the precision ranges from 93.44% to 100%, and the f1 score ranges from 96.61% to 100%. Thus, our RL model detects all the attacks, and there are some instances where attacks are detected incorrectly. Overall, the analyses show that the RL model is effective in turn-by-turn attack detection. In our follow-up study, we will explore the effectiveness of the RL-based approach for different types of sophisticated spoofing attack detection.

## ACKNOWLEDGMENTS

This material is based on a study partially supported by the National Science Foundation under Grant No. 2104999. Any opinions, findings, and conclusions, or recommendations expressed in this material are those of the author(s) and do not necessarily reflect the views of the National Science Foundation, and the U.S. Government assumes no liability for the contents or use thereof.

## AUTHOR CONTRIBUTIONS

The authors confirm contribution to the paper as follows: study conception and design: S. Dasgupta, T. Ghosh, and M. Rahman; data collection: S. Dasgupta and M. Rahman; interpretation of results: S. Dasgupta, T. Ghosh, and M. Rahman; draft manuscript preparation: S. Dasgupta, T. Ghosh, and M. Rahman. All authors reviewed the results and approved the final version of the manuscript.